\documentclass[aps,prl,twocolumn,superscriptaddress]{revtex4-2}
\usepackage{amssymb,graphicx,color}
\usepackage[intlimits]{amsmath}
\usepackage[english]{babel}
\usepackage[colorlinks]{hyperref}
\hypersetup{
colorlinks=true,
linkcolor=blue,
filecolor=blue,
urlcolor=blue,
citecolor=blue
}
\pdfoutput=1
\graphicspath{{figures/}} 
\usepackage[normalem]{ulem}
\usepackage{comment}
\usepackage{ragged2e}
\usepackage{enumerate}
\usepackage[dvipsnames]{xcolor}
\usepackage{dsfont}
\newcommand{\prj}[1]{\mathcal{P}_{|{#1}\rangle}}

\begin{document}

\title{Anomalously fast transport in non-integrable lattice gauge theories}

\author{Devendra Singh Bhakuni}
\affiliation{The Abdus Salam International Centre for Theoretical Physics (ICTP), Strada Costiera 11, 34151 Trieste, Italy}
\author{Roberto Verdel}
\affiliation{The Abdus Salam International Centre for Theoretical Physics (ICTP), Strada Costiera 11, 34151 Trieste, Italy}
\author{Jean-Yves Desaules}
\affiliation{Institute of Science and Technology Austria (ISTA), Am Campus 1, 3400 Klosterneuburg, Austria}
\author{ Maksym Serbyn}
\affiliation{Institute of Science and Technology Austria (ISTA), Am Campus 1, 3400 Klosterneuburg, Austria}
\author{ Marko Ljubotina}
\affiliation{Physics Department, Technical University of Munich,
TUM School of Natural Sciences, James-Franck-Str. 1, 85748 Garching, Germany}
\affiliation{Munich Center for Quantum Science and Technology (MCQST), Schellingstr. 4, M\"unchen 80799, Germany}
\author{Marcello Dalmonte}
\affiliation{The Abdus Salam International Centre for Theoretical Physics (ICTP), Strada Costiera 11, 34151 Trieste, Italy}

\begin{abstract}	

Kinetic constraints are generally expected to slow down dynamics in many-body systems, obstructing or even completely suppressing transport of conserved charges. Here, we show how gauge theories can defy this wisdom by yielding constrained models with \emph{faster-than-diffusive} dynamics. We first show how, upon integrating out the gauge fields, one-dimensional U(1) lattice gauge theories are exactly mapped onto XX models with non-local constraints. This new class of kinetically constrained models interpolates between free theories and highly constrained local fermionic models. We find that energy transport is superdiffusive over a broad parameter regime. Even more drastically, spin transport exhibits ballistic behavior, albeit with anomalous finite-volume properties as a consequence of gauge invariance. Our findings are relevant to current efforts in quantum simulations of gauge-theory dynamics and anomalous hydrodynamics in closed quantum many-body systems.
\end{abstract}

\maketitle

\textit{Introduction---}
Understanding the emergence of hydrodynamic behavior and the transport of conserved quantities in quantum many-body systems has been a central focus in recent times~\cite{bertini2016transport,bertini2021finite,castro2016emergent,denardis2020universality,denardis2021stability,doyon2017large,doyon2025generalized,abanin2019many,annurev_Huse,polkovnikov2011noneq,znidaric2011spin,Sommer2011universal,capizzi2025hydrodynamics,lux2014hydrodynamic,singh2021subdiffusion}. Locally interacting lattice field theories are expected to exhibit diffusive transport, with deviations arising in cases involving integrability~\cite{ljubotina2019kardar,denardis2020universality,denardis2021stability}, disorder~\cite{barlev2015absence,agarwal2015anomalous,Bar_Lev_2017}, kinetic constraints and higher-order conservation laws~\cite{feldmeier2020anomalous,gopalakrishnan2019kinetic,mccarthy2025subdiffusive,morningstar2020kinetically,moudgalya2021spectral,gromov2020fracton,richter2022anomalous,iaconis2021multipole,roy2020strong}, or finely tuned conditions~\cite{wang2025superdiffusive,znidaric2024superdiffusive}. Beyond these exceptions, identifying instances of locally interacting models that defy the generic expectation of diffusive transport remains an important open quest.

In this work, we show how gauge symmetries can induce faster-than-diffusive transport of conserved quantities in locally interacting systems. Concretely, we study transport properties of U(1) lattice gauge theories (LGTs) in $(1+1)$-dimension~\cite{kogut1975hamiltonian,wiese2013ultracold}, describing the coupling of fermions to gauge fields. As a key starting point, we analytically establish a duality between such LGTs and a new class of kinetically constrained models.
This formulation features highly non-local constraints resulting from the U(1) local symmetry. While such constraints may, at first glance, be expected to give rise to slower-than-diffusive transport~\cite{roy2020strong,smith2017absence,brenes2018many,Giudici2020breakdown, PhysRevLett.126.130401,Jeyaretnam2025}, here, the opposite happens: both energy and spin transport are faster than diffusive, with the latter being, in fact, ballistic. Remarkably, this occurs despite the system being non-integrable. 

Our findings are highly relevant to the field of quantum simulation of gauge theories, where recent experimental breakthroughs---both in digital and analog quantum simulations---have demonstrated a range of phenomena related to high-energy physics~\cite{Martinez2016,Bernien2017,zhou2021thermalization,farrell2024quantumsimulationshadrondynamics,gonzalezcuadra2024observationstringbreaking2,zhu2024probingfalsevacuumdecay,xiang2025realtimescatteringfreezeoutdynamics,schuhmacher2025observationhadronscatteringlattice,farrell2024scalable}, ubiquitously utilizing the U(1) quantum link models considered here~\cite{wiese2013ultracold,Dalmonte2016lattice,QLink1,QLink2,CHANDRASEKHARAN1997455,Banerjee2012,meglio2024quantum,pichler2016real,Banuls2020,bauer2023quantum}. In this context, our work offers a pathway for exploring hydrodynamic behavior differing from the traditional paradigm in models lacking local symmetries.

\begin{figure*}[t]
    \includegraphics[width=\linewidth]{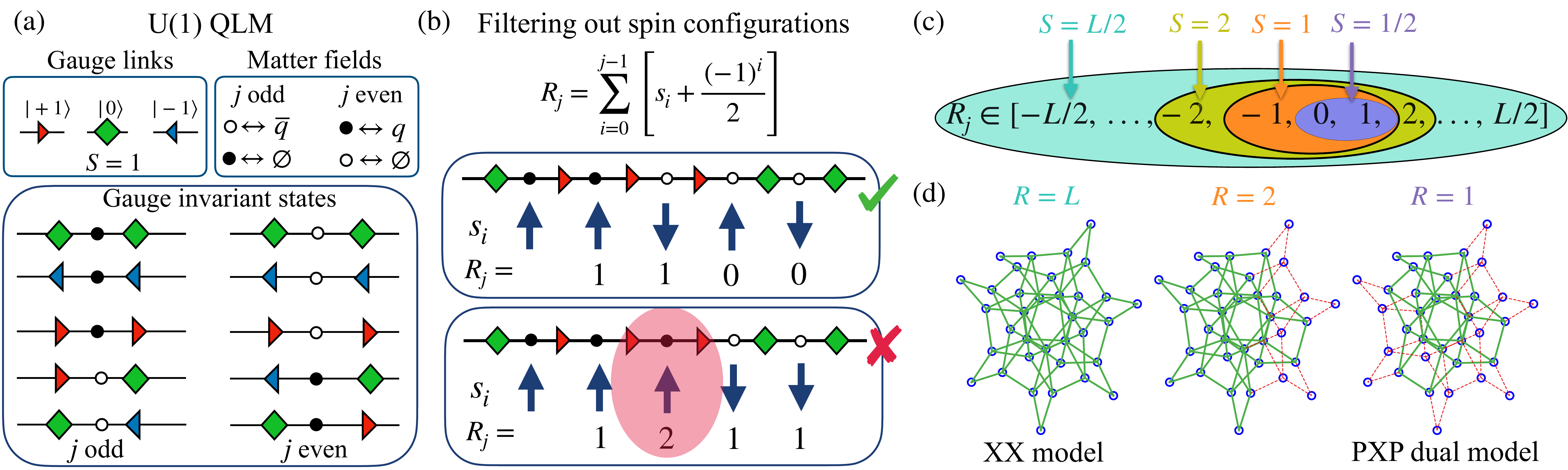}
\caption{ Schematic of the mapping between spin-$S$ QLM and the constrained XX models. An illustration for the $S=1$ QLM is shown in (a) and (b), which display the Hilbert space structure together with the convention for staggered fermions, and examples of allowed and discarded states according to Gauss's law, respectively. In this case, the latter restricts the values of the nonlocal constraint variable $R_j$ (where $s_i=\sigma^{z}_{i}/2$) to only $\{-1,0,1\}$. Through a Jordan-Wigner transformation of the matter fields, the valid configurations allow us to directly read off the basis states that span the space of a concomitant constrained spin model. (c) Generalization to the spin-$S$ case. The different ellipses represent the allowed values of $R_j$ for each value of $S$, defining a `constraint radius' $R=2S$ that gives rise to distinct constrained models. (d) Fock space visualization of the constraints: starting from the XX model ($R=L$), we discard the links to the nodes ruled out by the allowed values of $R_j$ [see  Eq.~\eqref{eq:11}] to obtain the constrained model for a given $R<L$. Green solid lines represent `active' links, while red dashed lines are `disconnected' links due to the gauge symmetry-induced constraint.} 
\label{fig:1}
\end{figure*}
\textit{Duality between U(1) LGTs and constrained XX models---}We consider the $(1+1)$--dimensional U(1) spin-$S$ quantum link model (QLM) describing the interactions of the matter fields mediated by the gauge fields~\cite{Banerjee2012}. The Hamiltonian reads:
\begin{align}
    \label{eq:1}
    H=&-w\sum_{j=0}^{L-2} (\psi^\dagger_j U_{j, j+1} \psi_{j+1} + \mathrm{H.c.}) \nonumber \\
    + &\  m \sum_{j=0}^{L-1} (-1)^j \psi^\dagger_j\psi_j +\frac{g^2}{2} \sum_{j=0}^{L} (E_{j-1,j})^2.
\end{align}
Here, $w$ is a hopping parameter for fermionic particles, and $g$ is the electric coupling strength. The matter field operator $\psi^\dagger_j$ ($\psi_j$) creates (annihilates) fermions at site $j$ with mass $m$. We adopt the Kogut-Susskind staggered fermion formulation~\cite{kogut1975hamiltonian}, where an occupied site corresponds to the vacuum on odd sites and to a quark $q$ on even sites, while an empty site corresponds to the vacuum on even sites and to an antiquark $\bar{q}$ on odd sites. The link variables $U_{j,j+1}$, defined between sites $j$ and $j+1$, are represented by spin operators of the spin-$S$ representation. Identifying $U_{j,j+1}=S^{+}_{j,j+1}$ and $U^\dagger_{j,j+1}=S^{-}_{j,j+1}$ as spin raising and lowering operators, and defining the electric field as $E_{j,j+1}=S^{z}_{j,j+1}$, ensures that the canonical commutation relations $[E_{j,j+1}, U_{j,j+1}] = U_{j,j+1}$ are satisfied~\cite{CHANDRASEKHARAN1997455}. In what follows, we restrict ourselves to the case $m=g=0$.

The U(1) gauge transformation generators are given by $G_j= E_{j,j+1} - E_{j-1,j} - Q_j$, where $Q_j = \psi^\dagger_j\psi_j  + \frac{(-1)^j-1}{2}$ is the charge of the fermions. The local gauge symmetry implies that $[H,G_j]=0$, such that the Hamiltonian $H$ does not mix eigenstates of $G_j$ with different eigenvalues. Here, we only consider the subspace defined by $G_j |\Psi\rangle=0, \; \forall j$, where Gauss's law links the difference in spin-$z$ projections to the local charge:  
\begin{equation}
    \label{eq:2}
   E_{j,j+1} - E_{j-1,j} \equiv S^{z}_{j,j+1} - S^{z}_{j-1,j}  = Q_j.
\end{equation}
Traditionally, in one dimension matter fields are eliminated via Gauss's law followed by  spin inversion on odd links, yielding the Hamiltonian~\cite{Surace2020lattice,desaules2023prominent,desaules2023weak}:  
$H_{G} = J \hat{\mathcal{P}} \left(\sum_{j} S^{x}_{j,j+1}\right) \hat{\mathcal{P}}$,  
where the projector $\hat{\mathcal{P}}$ enforces the constraint $S_{j,j+1}^{z} + S_{j-1,j}^{z} \in \{0, -1\}$.

In this work, we follow a different route and derive a class of constrained models by eliminating the gauge fields instead (see Fig.~\ref{fig:1}). Due to local gauge symmetry, the allowed charge configurations (with gauge fields fixed via Gauss's law) always satisfy $\Delta E_j \equiv E_{j-1,j}-E_\mathrm{left} = \sum_{i=0}^{j-1} Q_i  \in \{ -2S, -2S+1,\dots, 2S\}, \; \forall j>0$, where $E_\mathrm{left}\equiv E_{-1, 0}$ is the electric field on the left-end of the chain and $Q_i$ are the eigenvalues of the local charge operator (Fig.~\ref{fig:1}a). This arises from the fact that, in a spin-$S$ representation, the absolute change of the electric field between adjacent sites is at most $2S$, constraining the amount of charge that can be accumulated over a finite region of space. Furthermore, we work with boundary conditions where we fix the first and last electric field variables and set them to have the same value (one among the $2S+1$ possible values in $\{ -S, -S+1,..., S-1, S \}$). Specifically, we set them to the minimum possible absolute value of the field for integer spins, i.e., $E_\mathrm{left} =E_\mathrm{right}=0$, and to $E_\mathrm{left} =E_\mathrm{right}=-1/2$ for half-integer spins. This choice of boundary conditions also sets the net charge to be identically zero for an even number of sites. Other choices of boundary conditions give rise to the same higher-spin constrained model with smaller system sizes. 

Imposing the boundary conditions, and using $\psi_j^\dagger \psi_j =(\sigma^z_j+1)/2$, we define a constraint variable $R_j$ in the gauge allowed sector as (Fig.~\ref{fig:1}b-c)
\begin{equation}
    \label{eq:11}
 R_j \equiv \sum_{i=0}^{j-1} \Bigg(  \frac{\sigma^z_i}{2}  + \frac{(-1)^i}{2}\Bigg) \in\{-\Delta, -\Delta+1,\cdots, -\Delta+R\};
\end{equation}
$\forall \ j>0$, where $\sigma^z_i$ are the eigenvalues of the Pauli-$z$ operators and $\Delta=S$ for integer $S$, and $\Delta=S-1/2$ for half-integer $S$. The parameter $R=2S$ allows to tune the radius of the constraints and obtain different classes of constrained models. Denoting by $\mathcal{\bar{P}}$ the projector onto the resulting relevant subspace in the fermionic Hilbert space, and further using the Jordan-Wigner transformation~\cite{JordanWigner}: $\psi_j^\dagger= e^{i \pi \sum_{k<j} (\sigma^z_k+1)/2}\sigma_j^+, \
    \psi_j= e^{-i \pi \sum_{k<j} (\sigma^z_k+1)/2}\sigma_j^-$, the QLM can be recast into a spin-1/2 model with a non-local constraint
\begin{equation}
    \label{eq:8}
    H=\mathcal{\bar{P}} \Bigg(-w\sum_{j=0}^{L-1} (\sigma^+_j \sigma^-_{j+1} + \mathrm{H.c.}) \Bigg) \mathcal{\bar{P}},
\end{equation}
which is that of a constraint XX spin chain~\footnote{For a finite mass $m$, the constrained XX model gets an additional staggered longitudinal field term.} defined in the sector where the total magnetization $M=\sum_{i=0}^{L-1} \sigma^z_i$ takes the value $M=0$ or $M=-1$ depending on whether $L$ is even or odd, respectively~\footnote{If the labeling of lattice sites starts from an odd number, then for odd $L$ the constrained XX model is defined in the sector with total magnetization $M=1$ due to the convention for staggered fermions used in this work}. For $R=L$ (or $S=L/2$), which corresponds to an infinite spin representation for a finite system, the QLM reduces to the Schwinger model, which, after eliminating the gauge fields, takes the form of an unconstrained XX chain~\cite{PhysRevD.56.55, Muschik_2017}. For $R=1$, on the other hand, we obtain a highly constrained model that is dual to the PXP Hamiltonian~\cite{Surace2020lattice,PhysRevB.69.075106,TurnerNature,Lesanovsky2012}; see Fig.~\ref{fig:1}. 

The constraints imposed by gauge invariance can also be interpreted as disrupting the connectivity of the Fock-space graph (see schematic Fig.~\ref{fig:1}d). As mentioned above, this has the remarkable effect of yielding fast dynamics, in contrast with other ways of disrupting the Fock graph connectivity that have been linked to non-ergodicity and slow dynamics~\cite{roy2020strong}.

It is also important to note that while the Hilbert spaces of the matter-integrated QLMs and the constrained XX models [Eq.~\eqref{eq:8}] are identical for different values of $R$, their Hamiltonians differ in structure if $R>2$. In the matter-integrated QLM, the Hamiltonian contains non-uniform matrix elements due to higher-spin operators, whereas in the constrained XX models, these matrix elements remain uniform, making them distinct kinetically constrained models for $R>2$. We note that for both conventions we recover the lattice Schwinger model in the limit $S\to \infty$. In the following, we analyze the spectral and dynamical properties of the constrained XX models for different constraint radii $R$.

\textit{Spectral analysis---}In the unconstrained case ($R = L$), the XX model is exactly solvable and integrable. To determine whether the introduction of constraints breaks integrability, we examine the level statistics using the gap ratio, defined as $r_n = \min\left(\delta_{n}/\delta_{n+1}, \delta_{n+1}/\delta_{n}\right)$,
where $\delta_n$ is the energy gap between the $n^\text{th}$ and $(n-1)^\text{th}$ eigenvalues. The distribution of gap ratios, $P(r)$, serves as a key diagnostic. In a non-integrable system with a real Hamiltonian, $\delta_n$ follow the Wigner-Dyson distribution for the Gaussian Orthogonal Ensemble (GOE) which leads to $P(r) = 27(r + r^2)/4(1 + r + r^2)^{5/2}$. Meanwhile, in integrable systems $\delta_n$ exhibit Poisson statistics and thus $P(r) = 2/(1 + r)^2$~\cite{Atas2013levels,haakequantum,mehta2004random}.  

\begin{figure}[t]
    \includegraphics[width=\linewidth]{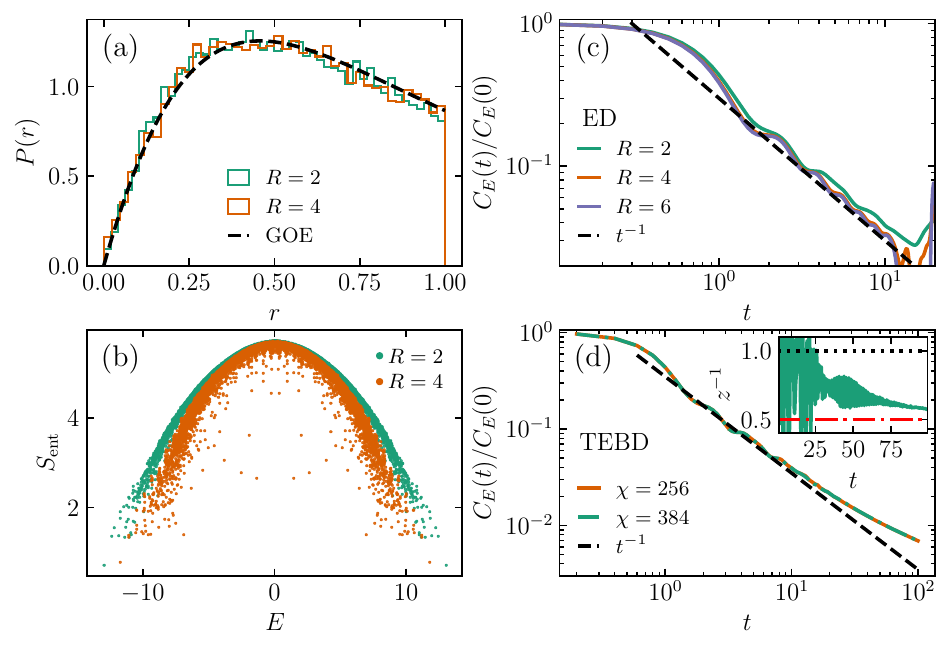}
\caption{(a,b) Spectral properties: (a) The gap-ratio distribution $P(r)$ for different values of constraint radii $R=2,4$ for $L=21,19$, respectively. The dashed line corresponds to the GOE prediction. (b) The half-chain entanglement entropy $S_\mathrm{ent}$ of each eigenstate as a function of energy. (c,d) Energy-energy autocorrelation function $C_{E}(t)$ using (c) exact-diagonalization for various constraint radii $R$ and $L=27$, and (d) TEBD simulations for $R=2$ and two different bond-dimensions $\chi=256,384$. The inset in (d) shows the inverse of the dynamical exponent $z^{-1}$.}
\label{fig:2}
\end{figure}

Using exact diagonalization (resolving reflection symmetry) for system sizes $L =21, 19$ with constraint radii $R = 2, 4$ respectively, we compute the gap-ratio distribution, shown in Fig.~\ref{fig:2}(a). For all $R$ values considered, $P(r)$ agrees well with the GOE predictions, indicating that the constraints break integrability. For larger constraint radii and for the accessible system sizes, deviations from the GOE predictions may occur, due to the chain being relatively short with respect to the range of the constraint.   

Next, we study the half-chain entanglement entropy of the eigenstates, defined as $S_\mathrm{ent} = -\text{Tr}[\rho_A \ln \rho_A]$, where $\rho_A = \text{Tr}_B[ \rho]$ is the reduced density matrix of subsystem $A$, obtained by tracing out the complementary subsystem $B$ from the pure state density matrix $\rho = |\psi\rangle\langle \psi|$. For non-integrable systems, the entanglement entropy typically exhibits volume-law scaling with system size and approaches the thermal value. As shown in Fig.~\ref{fig:2}(b), the eigenstate entanglement entropy for $R=2,4$ varies smoothly with energy. It approaches thermal values--particularly the Page value near zero energy density~\cite{Page1993Entropy}-- providing further evidence of non-integrability. 
For the matter-integrated QLMs with large spin $S$ representation of the gauge fields, a similar and consistent observation has been made in Refs.~\cite{desaules2023weak,desaules2023prominent}. These analyses suggest standard diffusion of the conserved quantities, namely, the energy and the spin. However, as we demonstrate below, this expectation does not hold.

\textit{Dynamical analysis: Energy transport---} We now turn to the dynamical properties and first focus on the nature of energy transport at infinite temperature by analyzing the energy-energy autocorrelation function 
\begin{equation}
    C_{E}(t) = \langle \mathcal{H}_{L/2}(t)\mathcal{H}_{L/2}(0)\rangle_{\infty} - \langle \mathcal{H}_{L/2}(t)\rangle_{\infty} \langle\mathcal{H}_{L/2}(0)\rangle_{\infty},
\end{equation}
where $\langle \cdot \rangle_\infty = \mathrm{Tr}[\cdot \rho_\infty]$, with $\rho_\infty$ being the infinite temperature density matrix, and $\mathcal{H}_{L/2} = -w(\sigma^+_{L/2} \sigma^-_{L/2+1} + \sigma^+_{L/2 +1} \sigma^-_{L/2})$ the energy density operator at the central bond. 
The energy-energy autocorrelation function decays as $C_{E}(t) \sim t^{-1/z}$ with the dynamical exponents $z$ characterizing the different transport regimes. A dynamical exponent of $z = 1$ signifies ballistic transport and $z = 2$ corresponds to diffusion, whereas super-diffusion (sub-diffusion) is characterized by $z < 2 (z > 2)$.

Using exact diagonalization (ED), we compute the dynamics of the energy-energy autocorrelation function $C_E(t)$ utilizing the dynamical typicality arguments~\cite{elsayed2013regression,bartsch2009dynamical,steinigeweg2014spin}
averaging $C_{E}(t)$ over $40$ random states. The results for constraint radii $R=2,4,6$ in a system of size $L=27$ are shown in Fig.~\ref{fig:2}(c).
Surprisingly, we observe clear ballistic transport with $z=1$ for all constraint radii, as opposed to the expected diffusive behavior. Interestingly, for $R>2$, the energy transport curves are nearly indistinguishable and closely match those of the unconstrained XX model within the system sizes and time scales considered---even though the constraints act non-perturbatively. This behavior is in stark contrast to the super-diffusive transport reported for the PXP ($R=1$ constrained) model~\cite{ljubotina2023superdiffusive,yupeng2024superdiffusive}.

To go beyond the time scales accessible with ED, we further perform tensor network simulations~\cite{schollwock2011density,xiang2023density,collura2024tensor} (see End Matter for details) using the $S=1$ QLM, which is equivalent to the $R=2$ constrained XX model. Figure~\ref{fig:2}(b) presents the decay of $C_E(t)$ for $L=256$ and two different maximum bond dimensions, $\chi=256$ and $\chi=384$. At long times, the behavior deviates from the ballistic regime and exhibits superdiffusive scaling. The inset shows the inverse dynamical exponent $z^{-1}$, extracted via a logarithmic derivative of $C_E(t)$ with respect to time for $\chi=384$. Initially, for times $t \lesssim 25$, $z^{-1}$ remains close to $1$, consistent with ballistic spreading. Beyond this window, $z^{-1}$ decreases from $1$, still remaining above the diffusive value $0.5$, for the accessed system size and time scale. However, we cannot exclude the possibility that the system may ultimately exhibit diffusion. Nonetheless, these results suggest that gauge-invariance constraints give rise to faster-than-diffusive transport, in contrast to other kinetic constraints that typically induce subdiffusion or even localization. We expect a similar trend to hold for larger constraint radii.

 \begin{figure}[t]
    \centering
\includegraphics[width=\linewidth]{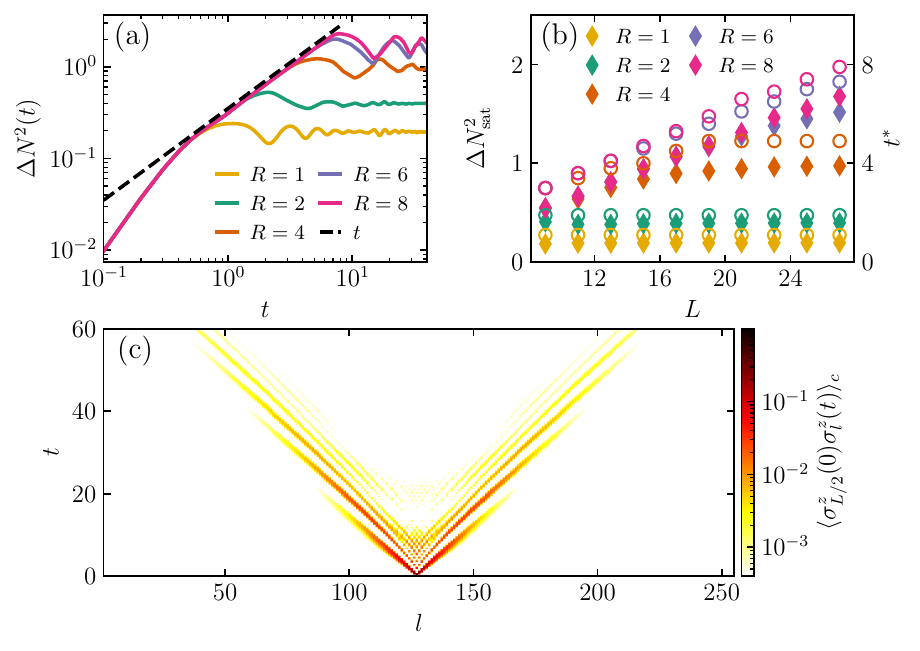}
    \caption{(a) Dynamics of particle number fluctuations $\Delta N^2(t)$ in a subsystem (of size $L/2$) for different constraint radii $R$ starting from a N\'eel state for $L=27$. (b) The saturation value at long time (filled diamonds) and the crossover time (empty circles) of the particle number fluctuations as a function of the system size $L$. (c) TEBD simulations: space-time profile of the connected spin-spin correlation function at infinite temperature for $R=2$ with $\chi=256$.}
    \label{fig:fig_3}
\end{figure}

\textit{Spin dynamics---}
The mapping introduced above allows for the immediate identification of a second conserved quantity, namely, the total magnetization $M=\sum_{i=0}^{L-1} \sigma^z_i$, enabling us to study particle number (or magnetization) fluctuations and spin transport. Starting with an initial state with no density variation (e.g., the N\'{e}el state), the subsystem particle number fluctuations are defined as
\begin{equation}
\label{eq:fluctuations}
\Delta N^2 (t) = \langle N^2(t) \rangle - \langle N(t) \rangle^2,
\end{equation}
where $N = \sum_{i\in \ell} n_i$ is the number of particles in a subsystem of size $\ell$.  Subsystem fluctuations are known to grow as $t^{1/2}$ for non-integrable systems, while for the XX model, they grow linearly in time~\cite{Wienand2024,fujimoto_dynamical_2021,bhakuni2024dynamic,Fujimoto2025exact}. We plot the dynamics of $\Delta N^2(t)$ for different constraint radii $R=1,2,4,6,8$ in Fig.~\ref{fig:fig_3}(a). We observe that for all values of $R$, the fluctuations $\Delta N^2 (t)$ follow the linear growth similar to the unconstrained XX model, albeit with different saturation values. We further analyze the system-size dependence of the saturation value at long-time $\Delta N^ {2}_{\text{sat}}$, and the crossover time $t^{*}$---the time when saturation begins. As depicted in Fig.~\ref{fig:fig_3}(a, b), both the saturation value and crossover time scale linearly up to a system size and then saturate for $R>2$, while for $R=1,2$, they are almost system size independent. For \emph{conventional} ballistic dynamics, these two quantities should instead scale linearly with the system size and never saturate~\cite{fujimoto_dynamical_2021,Fujimoto2025exact}. A behavior that we expect to recover in the limit $R \to \infty$ as they reach their plateau for larger and larger $L$ as $R$ increases. 

For a finite $R$, the anomalous scaling with the system size can be explained by the emergence of additional conservation laws imposed by Gauss's law and by the truncation of the gauge fields. Focusing on $R=1$, we see that the square of the charge imbalance between any two sides of the chain, $Q_\mathrm{imb}=(\sum_{i\in \mathcal{L}} Q_i - \sum_{i\in \mathcal{R}} Q_i )^2 = 1$, is also conserved. In terms of the number operators this reads: $(\sum_{i\in \mathcal{L}} n_i - \sum_{i\in \mathcal{R}} n_i  + \phi)^2 = 1,$ where $\phi  = \frac{L-2\ell}{2} - \frac{(-1)^{\ell}}{2},$ with $\ell$ being the number of sites in the left subsystem. Equation~\eqref{eq:fluctuations} then reduces to $\Delta N^2(t) = (1- 2\langle N(t)\rangle - \tilde{\ell}_{e/o})/4$, where $\tilde{\ell}_{e/o} = \ell\ \text{or}\ \ell +1$, depending on whether $\ell$ is even or odd. Thus, $\Delta N^2(t)$ evolves similarly to the local observable $N(t)$, with higher-order density correlations reducible to linear combinations of lower-order ones. Similarly, for $R=2$, any $n$-th ($n>4$) correlation function depends on lower-order terms, suggesting a finite time scale beyond which correlations become system-size independent. 

Finally, we probe the spin transport by focusing on the infinite-temperature connected spin-spin correlation function $\langle \sigma_{L/2}^z(0) \sigma_{l}^z(t)\rangle_c=\langle \sigma_{L/2}^z(0) \sigma_{l}^z(t)\rangle_{\infty}-\langle \sigma_{L/2}^z(0)\rangle_{\infty} \langle\sigma_{l}^z(t)\rangle_{\infty}$, using tensor network simulations for the $R=2$ case. We plot the space-time profile of the spreading of the correlations in Fig.~\ref{fig:fig_3}(c), revealing a linear light-cone with two fronts moving ballistically in opposite directions from the middle of the chain. 
We note that in the middle of the light-cone, correlations decay faster than ballistically. It is possible that this phenomenon is due to the constraint, as it severely limits the accumulation of spin in any region of space. This effect of the constraint can be seen directly in Eq.~\eqref{eq:11} or by looking at the expression for the conserved magnetization in the QLM (see End Matter).
It is important to note that, unlike energy that shows a departure from the ballistic behavior for larger system sizes, spin excitations show robust ballistic spreading up to the largest accessible time.

\textit{Discussion---}We introduced a new class of kinetically constrained quantum models arising from gauge invariance and showed that gauge invariance can lead to \emph{faster-than-diffusive} transport of conserved quantities, even though these systems feature signatures of non-integrability. This is reminiscent of previously studied non-integrable systems where ballistic transport emerges in the XXZ model due to the presence of a single impurity~\cite{Brenes2018high,brenes2020eigenstate,znidaric2020weak}. However, in our case, the constraints induced by gauge invariance act non-perturbatively and globally. Yet, we surprisingly recover superdiffusive transport on anomalously long time scales. We emphasize that the superdiffusion we observe is a model specific feature. For example, the other commonly used generalization of the PXP model to higher spin (not immediately related to a gauge theory) shows clear diffusion of energy (see End Matter). 

Our work is particularly relevant for the quantum simulations of gauge theories in two distinct contexts. For digital quantum simulators, the most convenient way would be to directly work in the dual formulation and apply the constraint as interspersed layers over Trotterization. For analogue simulators, proposals for realizing QLMs with various spin representations have been put forward~\cite{Banerjee2012,osborne2023spinsmathrmu1quantumlink}. Given the rapid progress in the field, we expect our findings to be accessible with next-generation platforms. 

\textit{Acknowledgments---}
We thank Riccardo Andreoni, Wouter Buijsman, Mario Collura, John Goold, Zala Lenarčič, Cristiano Muzzi, Romain Vasseur, and Marko Žnidarič for discussions. 
M.\,L. acknowledges support by the Deutsche Forschungsgemeinschaft (DFG, German Research Foundation) under Germany’s Excellence Strategy – EXC-2111 – 390814868.
M.\,L., and M.\,S. acknowledge support by the European Research Council under the European Union’s Horizon 2020 research and innovation program (Grant Agreement No.~850899). 
J.-Y.\, D.~acknowledges funding from the European Union’s Horizon 2020 research and innovation programme under the Marie Sk\l odowska-Curie Grant Agreement No.~101034413.
M.\,L., J.-Y.\,D. and M.\,S. acknowledge support by the Erwin Schrödinger International Institute for Mathematics and Physics (ESI). 
M.\,D. was partly supported by the QUANTERA DYNAMITE PCI2022-132919, by the EU-Flagship programme Pasquans2, by the PNRR MUR project PE0000023-NQSTI, and the PRIN programme (project CoQuS). 
M.\,D. and R.\,V. were supported by the ERC Consolidator grant WaveNets. 
D.\,S.\,B. acknowledges the CINECA Grant under the ISCRA-C (HP10C66VX2) program.
The tensor network numerical simulations were performed with the ITensor library~\cite{itensor}. 

\bibliography{ref}

\clearpage

\section{End Matter}
\textit{Tensor network simulations---}
For the tensor network simulations, we consider the $S=1$ QLM. In order to access the infinite temperature correlation functions within the desired sector of the Hilbert space, we prepare an initial projected identity matrix product operator (MPO) $\mathds{1}_0=\mathcal{L}\mathcal{M}\dots\mathcal{M}\mathcal{R}$, with
\begin{equation}
    \begin{split}
        \mathcal{L}=&\begin{pmatrix}\prj{0-} & \prj{-0}+\prj{00} & \prj{-+} \end{pmatrix},\\
        \mathcal{M}=&\begin{pmatrix}\prj{0-}+\prj{+-} & \prj{00} & 0 \\ \prj{0-} & \prj{-0}+\prj{00} & \prj{-+} \\ 0 & \prj{-0} & \prj{-+}\end{pmatrix},\\
        \mathcal{R}=&\begin{pmatrix}\prj{0-}+\prj{00}+\prj{+-} \\ \prj{-0}+\prj{0-}+\prj{00} \\ \prj{-0}\end{pmatrix},
    \end{split}
\end{equation}
where we hybridized pairs of neighboring sites to reduce the computational cost of the simulations, and introduced local projectors $\mathcal{P}_{|{x}\rangle}=|{x}\rangle\langle{x}|$. 
Then, depending on whether we are computing spin-spin or energy-energy correlation functions, we act on $\mathds{1}_0$ with the corresponding operator density to obtain the initial projected MPO. 

\begin{figure*}[t]
    \includegraphics[width=\linewidth]{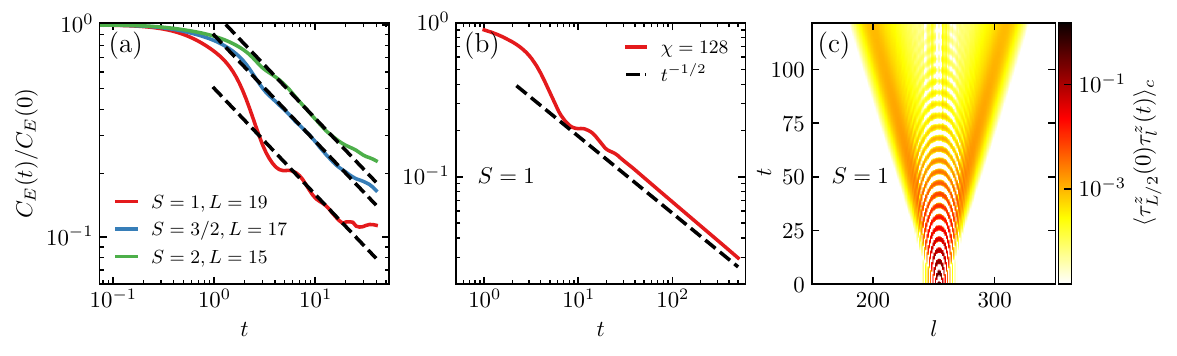}
\caption{Out-of-time connected correlation function in the spin-$S$ PXP model with OBC. (a)-(b) Energy-energy correlation function on the middle site of the chain computed using (a) ED and (b) tensor networks. In all cases, the decay quickly matches the expected diffusive behavior. The deviations from diffusion at late times are finite-size effects due to energy bouncing back at the edge of the system. (c) Full profile of the spin-spin correlation function. As in the QLM, spin transport is ballistic. However, the behavior inside the light cone is different, with persistent oscillations instead of rapid decay.}
\label{fig:PXP_S1}
\end{figure*}

This is then evolved in time using the time-evolving block decimation (TEBD) algorithm~\cite{schollwock2011density,xiang2023density,collura2024tensor}. 
To further minimize the computational resources, we fix the time step to $\delta t=0.2$ and use a fourth-order Trotter decomposition. 
We note that even this relatively large time step leads to errors that are smaller than those resulting from truncation, which we gauge by comparing the results obtained at several different bond dimensions. 

A similar procedure is used for the spin-1 PXP model, although the initial MPO and operators are modified accordingly. 

\textit{Magnetization in the QLM---}
In the main text, we have studied two conserved quantities of the constrained XX model: energy and spin. While it is clear that energy is also conserved in the QLM, this model does not have any apparent U(1) symmetry. The first step to solving this conundrum is to notice that even in the XX formulation, we restrict to a single magnetization sector. So in both cases, the Hilbert space is fully connected in the computational basis, without any splitting of U(1) sectors. If we then perform the mapping of the XX spins $\sigma^z_j$ in terms of QLM spins $S^z_{j,j+1}$, we find that 
\begin{equation}\label{eq:Seff}
s_j \equiv \sigma^z_{j}/2=(-1)^j\left(S^z_{j-1,j}+S^z_{j,j+1}+\frac{1}{2}\right),
\end{equation}
with the QLM constraint $S_{j,j+1}^{z} + S_{j-1,j}^{z} \in \{0, -1\}$, guaranteeing that $s_j=\pm 1/2$.
The staggering $(-1)^j$ leads to cancellations between the different terms, and it is straightforward to see that the total magnetization in a finite region of space only depends on the value of the $S^z$ at its boundaries. In a large enough region, this means that the maximum and minimum values the magnetization can take scales with $R$ (or equivalently $S$) but is independent of the region size. For a region going from the left end of the chain to a site $j$, this restriction is in fact exactly the constraint in Eq.~\eqref{eq:11} of the main text. Finally, if we now take the full system, the total magnetization will only depend on the boundary electric fields $E_\mathrm{left}$ and $E_\mathrm{right}$, which are frozen, hence it is conserved. 

So while there is no discrepancy between the conserved quantities of the constrained XX model and of the QLM, in the latter, the equivalent of the conserved spin is quite unusual. It is conserved somewhat trivially and is not extensive. In fact, one can engineer many similar conserved quantities by using the same construction $(-1)^j\left(h_{j-1,j}+h_{j,j+1}+C\right)$. Nonetheless, we emphasize that this effective spin conservation is special because of the constraint. Indeed, due to the form of the constraint it becomes the well-defined magnetization of the free XX chain in the $S\to \infty$ limit.

\textit{Spin-S PXP model---}
While QLMs offer a natural generalization of the PXP model to higher spins, the model can also be extended in other ways. Notably, in Ref.~\cite{WenWeiMPS}, a spin-$S$ PXP model was proposed, with a similar constraint that a spin can have $S^z>-S$ only if both of its neighbors have $S^z=-S$. This already differs from the QLM constraint for spin-1, as in the QLM the allowed neighboring configurations are $-0$, $-+$, $0-$, $00$ and $+-$, while for the spin-1 PXP model they are $--$, $-0$, $-+$, $0-$ and $+-$. We note that in general the spin-$S$ QLM is much more constrained, with the quantum dimension converging towards 2 as $S\to \infty$~\cite{desaules2023prominent} (equivalent to that of the Wilsonian theory). Meanwhile, for the spin-$S$ PXP it scales as $\sqrt{2S}$ in the same limit~\cite{WenWeiMPS}. Nonetheless, the two models display similar ergodicity-breaking, as both host quantum many-body scars that can be witnessed by quenching from the generalized N\'eel state $|-S,+S, \cdots, -S +S\rangle$~\cite{WenWeiMPS,desaules2023weak,desaules2023prominent}.

The transport properties are quite different for the two generalizations, with the spin-$S$ PXP model showing diffusion already at shorter times for $S>1/2$, see Fig.~\ref{fig:PXP_S1}. One possible explanation for this difference comes from the way they map to constrained models for degrees of freedom placed on the sites (assuming, as before, that PXP/QLM spins are placed on the links between them). For the QLM case discussed in the main text, the mapping is to a fermionic (or equivalently spin-1/2) chain. For the spin-$S$ PXP model, we instead need to consider a spin-$S$ particle on each site. These new degrees of freedom are then mapped as $\tau^z_{j}=(-1)^j[S^z_{j-1,j}+S^z_{j,j+1}+S]$. Notice that the only difference with the QLM expression in Eq.~\eqref{eq:Seff} is that the $+1/2$ has been replaced by $+S$. While the QLM constraint implies $S^z_{j-1,j}+S^z_{j,j+1}\in\{-1,0\}$, the PXP constraint instead leads to $S^z_{j-1,j}+S^z_{j,j+1}\in \{-2S,-2S+1,\ldots 0\}$ and so $S^z_{j-1,j}+S^z_{j,j+1}+S\in \{-S,-S+1,\ldots S\}$. As in the QLM, the staggered structure then enforces conservation of that effective spin-$S$, as if changing an $S^z_{j,j+1}$ leads to an increase of $\tau^z_j$ it must decrease $\tau^z_{j+1}$ by the same amount (and vice-versa).

In the PXP case, the resulting Hamiltonian on the sites is a spin-$S$ XX model with constraints. This has two important consequences. First, for $S>1/2$, this model is non-integrable even without constraints. As such, no ballistic transport would be expected even if the constraint could be treated perturbatively. Second, this also means that the constraint does not weaken as $S\to \infty$, unlike in the QLM where the free XX model is recovered in that limit. These two facts likely explain why we do not see any trace of superdiffusion in the spin-$S$ PXP model despite its apparent similarities to the spin-$S$ QLM.

One can ask whether this clear difference in energy transport between the two models is also present in spin transport. On Fig.~\ref{fig:PXP_S1} (c), we show the full profile of the out-of-time connected correlator $\langle \tau^z_{L/2}\tau^z_l\rangle_c $ for the spin-1 PXP model. The results are indicative of ballistic spin transport, very clearly differing from the energy diffusion in that model. This is somewhat similar to what happens in the QLM for $R=1$ and $R=2$, where spin transport is ballistic despite energy transport having $1/2<z^{-1}<1$. In both models, this difference in behavior between the two quantities is possibly due to spin not being extensive in this case and to the additional structure imposed on the spin by the constraint.

\end{document}